\documentclass[]{raa}            

\usepackage{graphicx,times}             
\usepackage{subfigure}
\usepackage{threeparttable}
\usepackage{float}
\usepackage[square,sort]{natbib}

\begin{document}

   \title{A machine learning method to separate cosmic ray electrons from protons from 10 to 100\,GeV using DAMPE data
}

   \volnopage{Vol.18 (2018) No.6, 000--000}      
   \setcounter{page}{1}          

   \author{Hao Zhao
      \inst{1,2}
   \and Wen-Xi Peng
      \inst{1}
   \and Huan-Yu Wang
      \inst{1,2}
   \and Rui Qiao
      \inst{1}
   \and Dong-Ya Guo
      \inst{1}
   \and Hong Xiao
      \inst{3}
   \and Zhao-Min Wang
      \inst{4,5}
   }

   \institute{Institute of High Energy Physics, Chinese Academy of Sciences,
             Beijing 100049, China {\it zhaohao@ihep.ac.cn, pengwx@ihep.ac.cn}
             \and
             University of Chinese Academy of Sciences,
             Beijing 100049, China \\
             \and
             Changzhou Institute of technology, Changzhou 213032, China
             \and
             Gran Sasso Science Institute (GSSI), Via Iacobucci 2, I-67100 L'Aquila, Italy
             \and
             Istituto Nazionale di Fisica Nucleare (INFN), Sezione di Lecce, Via per Arnesano, I-73100 Lecce, Italy
   }

   \date{Received~~2018 February 27; accepted~~2018~~March 17}

\abstract{DArk Matter Particle Explorer (DAMPE) is a general purpose high energy cosmic ray and gamma ray observatory, aiming to detect high energy electrons and gammas in the energy range 5\,GeV to 10\,TeV and hundreds of TeV for nuclei. This paper provides a method using machine learning to identify electrons and separate them from gammas, protons, helium and heavy nuclei with the DAMPE data acquired from 2016 January 1 to 2017 June 30, in energy range from 10 to 100\,GeV.
}

   \authorrunning{H.Zhao et al}            
   \titlerunning{Machine learning to separate CREs from protons}  

   \maketitle

%
%
\section{introduction}           
\label{sect:intro}

The high energy spectrum of cosmic rays electrons (CREs) has been investigated both on the ground and in space in recent years \citep{aguilar2014electron, picozza2007pamela, aharonian2008energy}, in order to gain knowledge on CREs' origin and propagation, and try to search for signals of dark matter annihilation \citep{pospelov2009astrophysical}. The direct detection of CREs in space is more accurate than ground detections since there is no atmosphere between the primary CREs and the detectors. Due to the small effective area of satellite experiments, the electrons flux could be measured up to a few TeV. A direct detection of electrons up to 10\,TeV may reveal some unknown physics related to dark matters or nearby sources \citep{kobayashi2004most,yin2013pulsar}.

The DArk Matter Particle Explorer (DAMPE) was successfully launched into a Sun-synchronous orbit at an altitude of 500\,km on 2015 December 17 from the Jiuquan Satellite Launch Center \citep{chang2017dark}. DAMPE mainly focuses on the detection of galactic cosmic rays (GCRs), the potential signal of dark matter annihilation and gamma-ray astronomy. It consists of four sub-detectors, a Plastic Scintillator strip Detector (PSD), a Silicon-Tungsten tracKer-converter (STK), a BGO (Bi4Ge3O12) imaging calorimeter and a NeUtron Detector (NUD). The PSD consists of two layers of scintillators to measure the charge of passing nuclei and as veto detector for gamma rays; the STK consists of 12 layers of single-sided silicon detectors.It can measure the trajectory and charge of charged particles, identify nuclei up to oxygen and the tungsten plates inside can convert photons into $e^+e^-$ pairs. The BGO calorimeter is made of 14 layers where each layer contains 22 BGO bars. It measures the energy profile of an electromagnetic shower induced by electrons, positrons or photons, or a hadronic shower induced by protons and nuclei. The NUD records the secondary neutrons produced in a BGO shower and contributes to the identification of electrons/protons.

It's a big challenge to get clean samples of electron from GCR data since the number of background protons is approximately $10^3$ times larger than electrons above hundreds of GeV. The traditional cuts-based method used in \cite{ambrosi2017direct} relies on a full understanding of the detector but does not take into account all the relations between the variables used and in this way could miss some hidden piece of information. The machine learning method has been widely used in data analysis of various physics experiments in the last decades and usually gives better results than the traditional cut-based method \citep{roe2005boosted,abdollahi2017cosmic}. In this paper we present an analysis using the machine learning method to separate electrons from background, mainly protons, in energy range from 10 to 100\,GeV.

The whole procedure used in the analysis is introduced in Section~\ref{sect:procedure}. The data analysis is presented in Section~\ref{sect:Pre-Analysis} and Section~\ref{sect:analysis} in detail. A discussion about the method is in Section~\ref{sect:discussion} and the conclusion given is in Section~\ref{sect:conclusion}.


\section{Analysis procedure}
\label{sect:procedure}
\begin{figure}[htbp]
   \centering
   \includegraphics[width=0.66\textwidth]{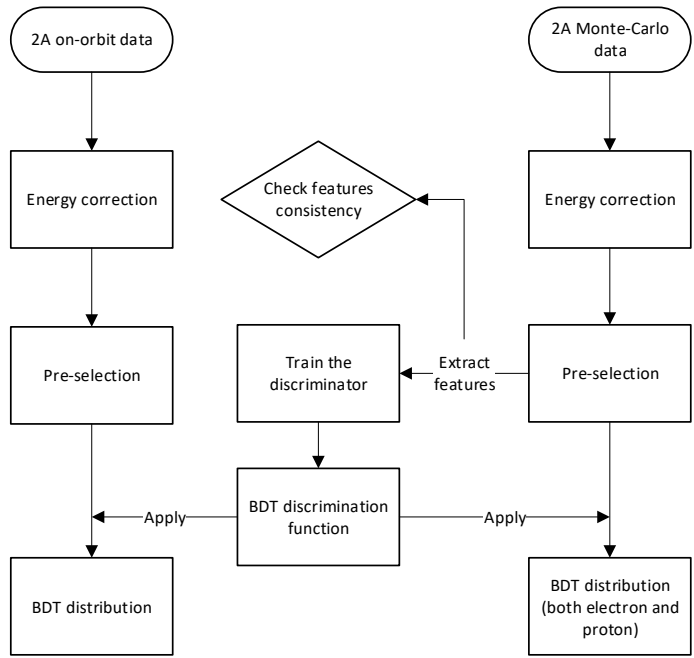}
   \caption{The flow chart of the machine learning analysis.}
   \label{fig:flowchart}
\end{figure}

The analysis starts with DAMPE 2A ROOT files (Monte Carlo (MC) data from version 5.3.0), which have already been properly calibrated and reconstructed \citep{zhang2016calibration}. A flow chart illustrating the machine learning method is shown in Figure~\ref{fig:flowchart}.

The whole process starts with an energy correction process to correct the energy of electron events (Sect.~\ref{sect:Energy-Correction}). Since there are lots of events with large incident angles, events in which the shower is partially recorded in the BGO or events from the bottom of the detector, the energies may not be precisely reconstructed. A pre-selection is applied to eliminate those events in both MC data and on-orbit data (Sect.~\ref{sect:pre-selection}). After the pre-selection, we extract some characteristic variables from BGO (which represents features of the shower) of all the surviving electrons and protons in the MC data(Sect.~\ref{sect:variables}), then feed the variables into the machine learning training algorithm to yield a "discriminator" to distinguish electrons from protons (Sect.~\ref{sect:machinelearning}).

\section{Data Pre-Selection}
\label{sect:Pre-Analysis}

\subsection{Energy Correction}
\label{sect:Energy-Correction}

\begin{figure}[htbp]
   \centering
   \includegraphics[width=0.66\textwidth]{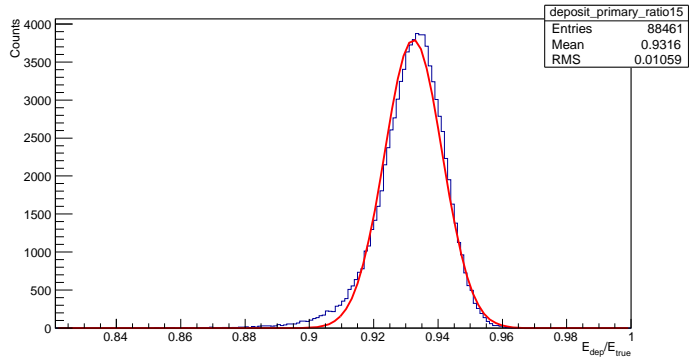}
   \caption{A gauss fit to the distribution of MC electrons' energy ratio $E_{dep}/E_{true}$ in deposited energy bin 86.6-100\,GeV, where $E_{dep}$ is the deposited energy of electrons and $E_{true}$ is the true energy. Fit result: mean=0.929; sigma=0.011.}
   \label{fig:energyratio}
\end{figure}

As the first step, the energy of an electron event needs to be corrected. The total energy of the incident electron may not be fully collected because some energies of the shower might be deposited in the supportive structure or leak from the side or bottom of the detector. So, it is necessary to correct the deposited energy from electrons to compute their original energy. We use the MC data to evaluate the deposited energy distribution for electrons. An example is shown in Figure~\ref{fig:energyratio}. The distribution of deposited energy/MC truth energy ratio approximates a Gaussian distribution.

\begin{figure}[htbp]
   \centering
   \includegraphics[width=0.66\textwidth]{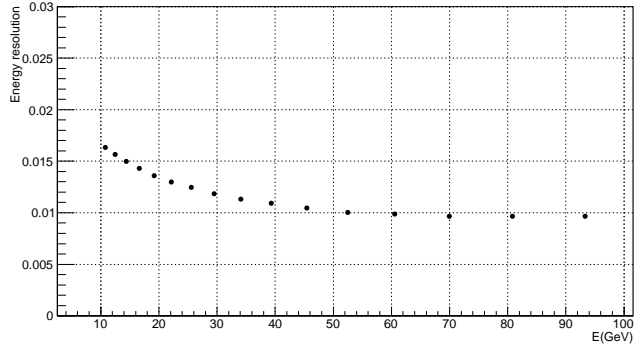}
   \caption{Energy resolution from the Gaussian fit of the MC electrons versus the deposited energy.}
   \label{fig:energyres}
\end{figure}

We fit data displayed in Figure~\ref{fig:energyratio} with a Gaussian distribution, and calculate the mean and sigma values. The mean value is taken as the ratio between the deposited energy and the MC truth energy of incident electrons for all the events in this energy bin. The electron energy resolution is based on the MC data from 10 to 100\,GeV, which is the sigma from the Gaussian fit, as shown in Figure~\ref{fig:energyres}. This resolution is well-matched with the beam test result, indicating a good configuration for the simulation \citep{zhang2016calibration}.

The energy correction process needs to be applied to all on-orbit data since we cannot tell if a given event results from an electron or not. Through this process, the energies of electrons are corrected but the energies of protons are also changed. Since our final objective is to get a clean electron sample and reconstruct the electron spectrum later, this is not a concern as long as the electron energy is precise.
\subsection{Pre-selection Cut}\label{sect:pre-selection}

After the energy correction, we need to select events whose energies and trajectories can be reconstructed accurately. We call this procedure 'pre-selection'. Several cuts are implemented to remove events that have poor energy reconstruction, both on MC and on-orbit data. Here are those cuts in order:

1. There is at least one \emph{Globtrack} \citep{chang2017dark} reconstructed successfully. The \emph{Globtracks} of an event are its STK tracks which match the BGO reconstructed track. Then we select the one with the smallest chi-square value from the track reconstruction for further analysis.

2. Use PSD reconstructed energy to eliminate He and heavier nuclei.

3. Use PSD reconstructed energy to eliminate photons.

4. \emph{Globtrack} goes through both the top and bottom layer of the BGO calorimeter.

5. The bar with the maximum deposited energy is not on the edge of the calorimeter.

6. The ratio between the total \emph{RMS} and \emph{HorizontalRMS} (defined in Sect.~\ref{sect:variables}) is larger than 15 to exclude particles incident from the flank of DAMPE.

7. The energy deposition ratio of the first BGO layer is less than 0.2. This cut tries to eliminate those events with large incident angles, but which cannot be removed by cut 4 since the reconstructed \emph{Globtrack} of these events is inaccurate.

\begin{figure}[htbp]
\centering
\begin{minipage}[b]{\textwidth}
\subfigure[]{
\includegraphics[width=.46\textwidth]{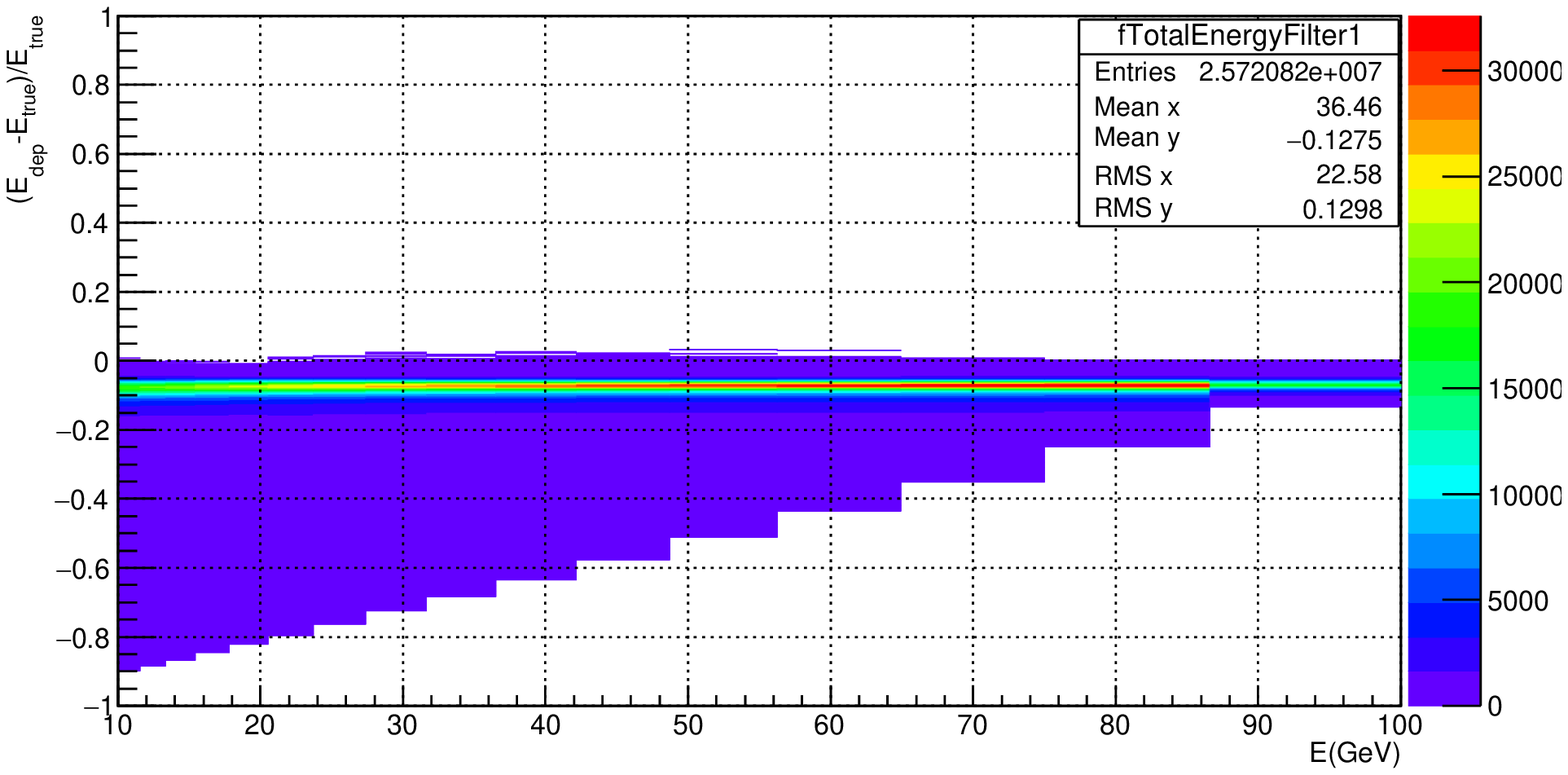}
}
\subfigure[]{
\includegraphics[width=.46\textwidth]{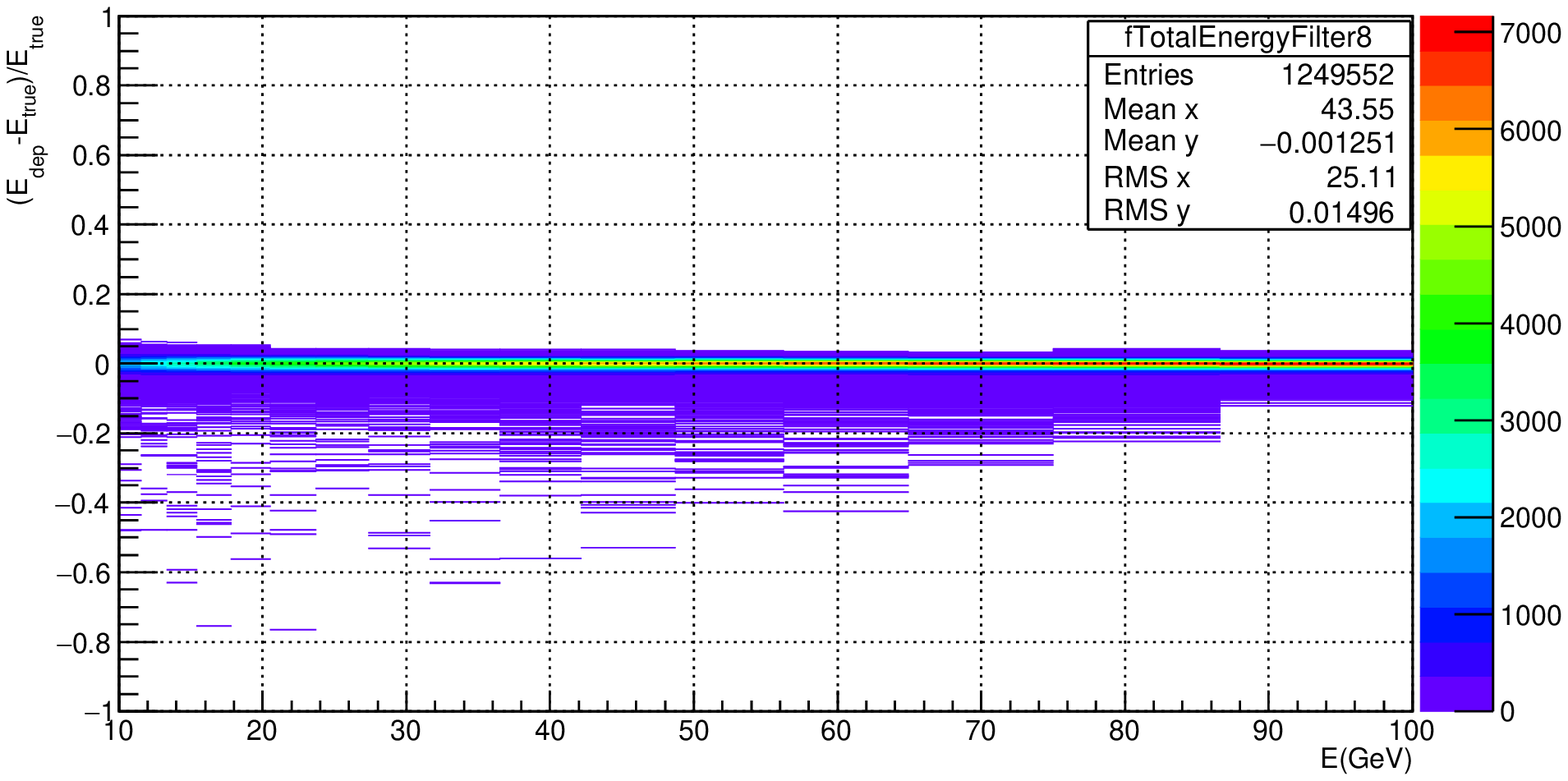}
}
\end{minipage}
 \caption{Deposited energy of events. (a) Before all cuts. (b) After all cuts. The x axis is the deposited energy. The y axis is $(E_{dep}-E_{true})/E_{true}$, where $E_{dep}$ is the deposited energy; $E_{true}$ is the true energy.}
 \label{fig:cut}
\end{figure}

Figure~\ref{fig:cut} shows the distribution of the deposited energy after all cuts. Those events with y-axis value close to 0 are fully deposited events which need to be preserved. It is obvious that most events with small deposited energy have been excluded by these cuts. After all cuts are applied, about 0.1\% of partially deposited events remain in the whole sample, which is acceptable considering the error from this effect is about 0.1\%.

Moreover, the pre-selection can depress the proton/electron ratio due to differing profiles of electromagnetic and hadronic showers. Table~\ref{table:cutresponse} shows the statistics of electrons and protons with these cuts based on the MC data. The bottom row shows the percentage of surviving particles compared to all particles. Consequently, the percentage of electrons is 3 times more than the percentage of protons, so the pre-selection has a background rejection power of 3 in the energy range 10-100\,GeV.

\begin{threeparttable}[htbp]
\begin{center}
\caption[]{\label{table:cutresponse}Electrons and Protons' Response to the Cuts. The percentage shown in the table is the ratio of surviving particles after the corresponding cut to the particles before this cut. }
 \begin{tabular}{lcccc}
  \hline\noalign{\smallskip}
  & \multicolumn{2}{c}{Electrons}   & \multicolumn{2}{c}{Protons}    \\
  \hline\noalign{\smallskip}
Cut 0  &  $2.57\times10^7$  & $100\%$  & $2.46\times10^7$ & $100\%$   \\
Cut 1  &  $1.39\times10^7$  & $54.1\%$ & $1.36\times10^7$ & $55.3\%$  \\
Cut 2  &  $6,426,500$       & $46.2\%$ & $4,488,086$        & $33.0\%$  \\
Cut 3  &  $3,613,769$       & $56.2\%$ & $1,526,301$        & $34.1\%$  \\
Cut 4  &  $1,488,657$       & $41.2\%$ & $690,336$        & $45.2\%$  \\
Cut 5  &  $1,419,133$       & $95.9\%$ & $600,799$        & $87.0\%$  \\
Cut 6  &  $1,305,428$       & $92.1\%$ & $488,117$        & $81.2\%$  \\
Cut 7  &  $1,249,552$       & $96.1\% (4.9\%^1)$& $464,585$  & $99.7\% (1.9\%)$ \\
  \noalign{\smallskip}\hline
\end{tabular}
\begin{tablenotes}
\footnotesize
\item[1] The value in brackets is the percentage of the surviving events of all cuts.
\end{tablenotes}
\end{center}
\end{threeparttable}


\section{Data analysis}
\label{sect:analysis}

We now should consider how to increase electron rejection power and separate them from the severe background events. As the BGO calorimeter has a much larger radiation length than the nuclear interaction length, an electromagnetic shower is typically thinner and smaller than the hadronic shower with the same deposited energy. In order to quantify the shape of a shower, we define some characteristic variables to describe its profile.

\subsection{Define Characteristic Variables}\label{sect:variables}

The characteristic variables are mainly based on the differences between electromagnetic showers and hadronic showers in the BGO calorimeter. These variables can be studied using the MC data and are listed in the following. The whole detector is defined in a three-dimension coordinate system. The orientation of positive z is vertical to the layer of BGO crystals pointing at the STK detector while x and y are defined in the plane that is parallel with the layer.

\subsubsection{Energy Deposition Fraction in Each Layer}

The variable \emph{$Efrac_i$} is simply defined as the energy deposit in the \emph{i-th} layer divided by the total deposit energy, where \emph{i} is the layer number from 0 to 13. Since a hadronic shower tends to extend more deeply than an electromagnetic shower in the BGO calorimeter, the energy deposition fraction of the last few layers has strong discriminating power.

\subsubsection{Root Mean Square}
The \emph{RMS} (Root Mean Square) of the deposited energies in the \emph{i-th} layer is defined as
\begin{equation}\label{equ:rms}
  RMS_i = \sum_{j=0}^{22} E_{ij}\times(d_{ij}-d_i^{cog})^2
\end{equation}
where $E_{ij}$ is the deposited energy of the \emph{j-th} bar in the \emph{i-th} layer, $d_{ij}$ is the one-dimensional coordinate of the \emph{j-th} bar in \emph{i-th} layer and $d_i^{cog}$ is the one-dimensional coordinate of the \emph{i-th} layer's center of gravity, defined as
\begin{equation}\label{equ:cogrms}
  d_i^{cog} = \sum_{j=0}^{22} E_{ij}\times d_{ij}/E_i
\end{equation}
Here $E_i$ is the deposited energy of the \emph{i-th} layer, so we can get a total of 14 \emph{RMS} values for BGO. Total \emph{RMS} used in Sect.~\ref{sect:pre-selection} is simply the summation of all 14 \emph{RMS} values. Based on the definition, the \emph{RMS} indicates the lateral development of the shower. The electromagnetic shower is expected to be thinner than the hadronic shower when the incident particle has the same energy, resulting in a smaller \emph{RMS} value.

\subsubsection{$RMS_r$ and $RMS_l$}
\emph{$RMS_r$} \citep{lixiangPhDThesis2016} is the root mean square where the center of gravity of energy deposit is along the \emph{Globtrack}. \emph{$RMS_r$} is defined as
\begin{equation}\label{equ:rmsr}
  RMS_r = \sqrt{\sum_{i=1}^{N} E_i\times D_i^2/E_{total}}
\end{equation}
where N represents all the triggered BGO bars, $E_{total}$ is the total deposited energy in BGO and $D_i$ is the distance between the \emph{i-th} BGO bar and the \emph{Globtrack}.

\emph{$RMS_l$} is similar to \emph{$RMS_r$} except that $D_i$ is the distance between the \emph{i-th} BGO bar and the center of gravity which is defined as
\begin{equation}\label{equ:cogrmsl}
  d_{cog} = \sum_{i=0}^{N} E_{i} \times d_{i}/E_{total}
\end{equation}
where $d_i$ is the three-dimensional coordinates of the \emph{i-th} BGO bar. These two variables are evaluated for the whole BGO detector. \emph{$RMS_r$} and \emph{$RMS_l$} also indicate the lateral development of the shower. These two values are less dependent on the structure of the detector since they are only calculated using the actual track of the incident particle.

\subsubsection{FValue}
\emph{$FValue_i$} of the \emph{i-th} layer is defined as
\begin{equation}\label{equ:fvalue}
  FValue_i = \sum_{j=0}^{22} E_{ij}\times(d_{ij}-d_i^{cog})^2\times Efraction_i
\end{equation}
where $Efraction_i$ is the fraction representing the deposited energy of the \emph{i-th} layer to the total deposited energy. Other parameters are the same as defined in Equation~\ref{equ:rms}. \emph{$FValue_i$} is a combination of \emph{$RMS$} and \emph{$Efrac_i$}, which show stronger distinguishing power than these two variables alone \citep{chang2008excess}.

\subsubsection{L factor}
\emph{L factor} \citep{lixiangPhDThesis2016} is a combination of $RMS_l$ and \emph{FValue}, defined as
\begin{equation}\label{equ:lfactor}
  L=(\frac{RMS_l}{60})^{1-\alpha}(\frac{F_{13}+F_{14}}{0.1})^\alpha
\end{equation}
where $F_{13}$ and $F_{14}$ are \emph{$FValue_{13}$} and \emph{$FValue_{14}$}, respectively; $\alpha$ is defined as:
\begin{equation}\label{equ:alpha}
\alpha=0.5+\frac{1}{\pi}arctan(5\log_{10}(\frac{E}{50GeV}))
\end{equation}
\emph{L factor} also considers the energy factor, which allows the variable to have steady performance over the whole energy range.

\subsubsection{Chi-square}
\cite{longo1975monte} present an equation to describe the electromagnetic shower as
\begin{equation}\label{equ:emshower}　　
  dE/dx=k_{norm}t^ae^{-bt}
\end{equation}
where \emph{a} and \emph{b} are coefficients, and $k_{norm}$ is defined as
\begin{equation}\label{equ:knorm}　
  k_{norm}=Eb^{a+1}/\Gamma(a+1)
\end{equation}
where $\Gamma$ is the Euler's Gamma Function. We use this equation to fit the shower and compute the associated \emph{Chi-square}. The \emph{Chi-square} from the fit is a good estimator to distinguish electromagnetic showers from hadronic showers. Obviously, the \emph{Chi-square} of an electromagnetic shower is smaller compared to the \emph{Chi-square} of a hadronic shower with the same deposited energy.

\subsubsection{tmax}
The variable \emph{tmax} is derived from Equation~\ref{equ:emshower} after the fit to the shower. It is defined as
\begin{equation}\label{equ:tmax}　
  tmax=a/b
\end{equation}
The variable \emph{tmax} also represents the radiation length where the energy loss $\frac{dE}{dx}$ become the largest.

\subsubsection{LongitudinalRMS}
\emph{LongitudinalRMS} is defined as:
\begin{equation}\label{equ:LongitudinalRMS}　
  LongitudinalRMS=\sum_{i=0}^{14}E_i\times(dz_i-dAll_{cog})
\end{equation}
where $dz_i$ is the z coordinate of the \emph{i-th} layer, $dAll_{cog}$ is the center of gravity for all layers, which is calculated as
\begin{equation}\label{equ:dcog}
  dAll_{cog}=\sum_{i=0}^{14}E_i \times d_i/E
\end{equation}
where \emph{E} is the total deposited energy. \emph{LongitudinalRMS} indicates the longitudinal development of the shower, which turns out to have strong discrimination capability as shown in Figure~\ref{fig:varidis}.

\emph{HorizontalRMS} used in Section~\ref{sect:pre-selection} is defined in the same way as in Equation~\ref{equ:LongitudinalRMS} except that $d_i$ is the x or y coordinate of the \emph{i-th} bar.

\subsection{Variable Comparison}\label{sect:comparison}

\begin{figure}[htbp]
\centering
\includegraphics[width=.66\textwidth]{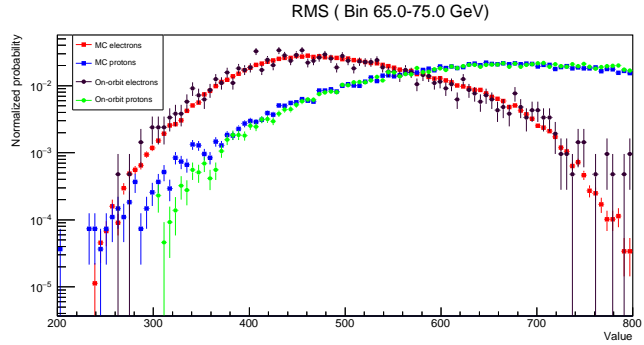}
 \caption{The MC and on-orbit data comparison of variable \emph{$RMS$}, only in 65.0-75.0\,GeV bin. Green: on-orbit proton; Black: on-orbit electron; Blue: MC proton; Red: MC electron.}
 \label{fig:comparison}
\end{figure}

Since the discriminator will be trained using the MC data, it is crucial to check the consistency between the MC data and the on-orbit data. First of all, we use 3 variables, \emph{FValue}, \emph{$L\_factor$} and \emph{RMS}, to select electrons and protons preliminarily from on-orbit data for variable comparison. Then all variables in every energy bin of the MC and the on-orbit data are compared. All the variables defined in Section~\ref{sect:variables} which show good consistency in every energy bin will be used in further analysis. An example of the comparison for variable (\emph{$RMS$}) is displayed in Figure~\ref{fig:comparison}. The match between MC and on-orbit data is good based on visual inspection in this energy bin. After passing the comparison test in all energy bins, this variable (\emph{$RMS$}) will be used for training.

\subsection{Machine Learning}
\label{sect:machinelearning}

We use the boosted decision trees (BDT) method from the toolkit TMVA \citep{hoecker2007tmva} to train the network. A decision tree is a binary tree structured classifier, in which each internal node is labeled with an input feature. The boosting of a decision tree extends this concept from one tree to several trees which form a forest. Boosting is a way of enhancing the classification and regression performance (and increasing the stability with respect to statistical fluctuations in the training sample) of typically weak multi-variant analysis (MVA) methods by sequentially applying an MVA algorithm to reweighted (boosted) versions of the training data and then taking a weighted majority vote of the sequence of MVA algorithms thus produced. It has been introduced to classification techniques in the early 1990s and in many cases this strategy has resulted in dramatic performance increases \citep{freund1999short,schapire2003boosting}.

\begin{figure}[htbp]
\centering
\includegraphics[width=0.66\textwidth]{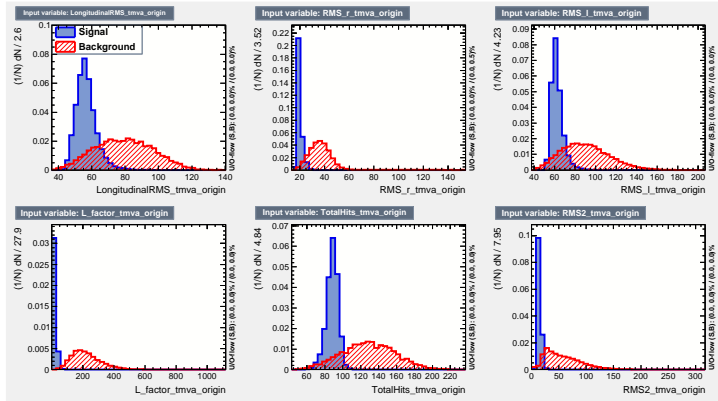}
 \caption{The variable distributions of \emph{LongitudinalRMS}, \emph{$RMS_r$}, \emph{$RMS_l$}, \emph{$L\_factor$}, \emph{TotalHits} and \emph{RMS} of the MC electrons and protons in energy bin 65.0-75.0\,GeV. Blue: electrons; red: protons. }
 \label{fig:varidis}
\end{figure}

MC electrons and protons passing pre-selection are used for the machine learning. Half of the sample is used as training sample and the other half as test sample. The values of the variables for every MC event, forming a vector, will be input into the training algorithm. The algorithm will generate a discriminator which can distinguish electrons from protons. Every energy bin in range 10-100\,GeV is trained separately. We show the training result for the 65.0-75.0\,GeV energy bin as an example. The distributions of some well distinguished variables are shown in Figure~\ref{fig:varidis}.

\begin{figure}[htbp]
   \centering
   \includegraphics[width=0.66\textwidth]{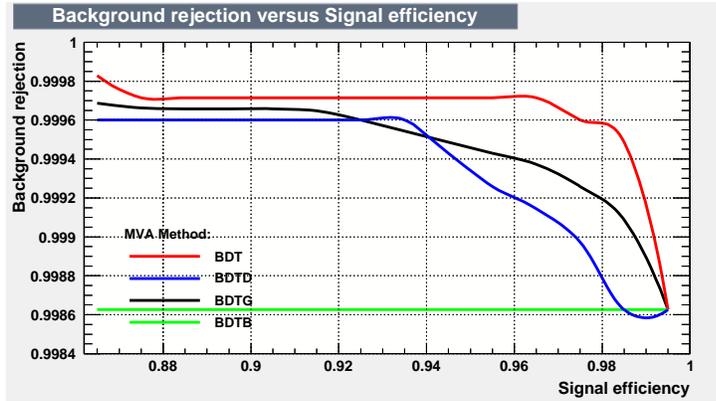}
   \caption{Receiver Operating Curve (ROC) in the energy bin 65.0-75.0\,GeV, where X axis is the signal selection efficiency and Y axis is the background rejection power. The results of the four different BDT methods are shown for comparison.}
   \label{fig:roc}
\end{figure}

The Receiver Operating Curve (ROC) \citep{bradley1997use} in the 65.0-75.0\,GeV energy bin that result from four BDT methods, namely BDT with adaptive boosting (BDT), BDT with gradient boosting (BDTG), BDT with bagging (BDTB), BDT with decorrelation and adaptive boosting (BDTD) \citep{hoecker2007tmva}, are shown in Figure~\ref{fig:roc}. The background rejection power of the BDT, the selected method, is about 0.9997 at signal selection efficiency of 0.9. Also, in a certain energy bin, the statistic corresponding to protons is mainly contributed by protons with higher energy \citep{zhang2015design}. Based on MC data, this so-called "hard suppression" has a suppression power of 9. With hard suppression of 9, the pre-selection cut's suppression power of 3 we mentioned before and the suppression power provided by BDT, this analysis gets total $3\times9\div(1-0.9997)=9.0\times10^4$ for background rejection power at 0.9 signal selection efficiency after pre-selection in the 65.0-75.0\,GeV energy bin. Similar background rejection powers are obtained in other energy bins of 10-100\,GeV.

\begin{figure}[htbp]
\centering
\includegraphics[width=.66\textwidth]{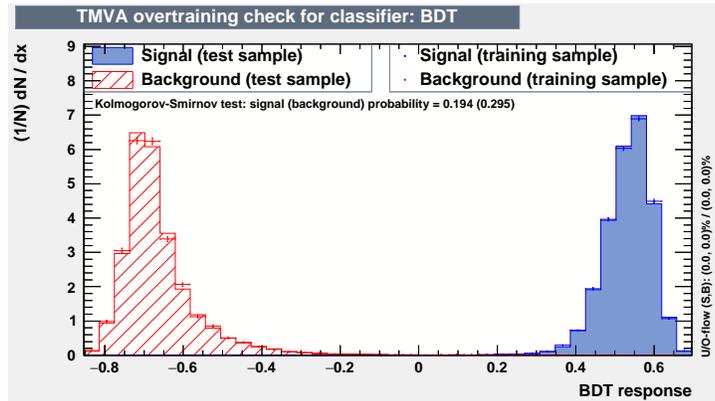}
 \caption{The output distributions of the BDT method from both training and test sample in energy bin 65.0-75.0\,GeV. The Kolmogorov-Smirnov test is performed to check overtraining.}
 \label{fig:bdtdis}
\end{figure}

Applying the discriminator to both training and testing sample, the BDT response distributions of both electrons and protons in the 65.0-75.0\,GeV energy bin are obtained as shown in Figure~\ref{fig:bdtdis}. These distributions are considered as templates in the template fit to calculate the counts of electrons and protons, a topic which is not the content of this paper. Also an overtraining check is performed to check if the distributions of training sample and test sample are consistent. The signal value of 0.194 and the background value of 0.295 indicate no overtraining is present.

\section{Discussion}
\label{sect:discussion}

There are also many other machine learning methods such as Support Vector Machine (SVM), Artificial Neutral Network (ANN) and K-Nearest Neighbor (KNN) available in TMVA package. It is crucial to choose appropriate and strong distinguishing characteristic variables in order to have higher background rejection power. It is worth pointing out some of our BDT training settings: number of trees is 1000; max depth of a tree is 3; number of grid points in variable range used in finding optimal cut in node splitting is 20. We have altered some of these parameters in order to get better training result. Since the BDT result we got is almost perfect, the tuning did not yield much difference in this case.

In addition, the field of deep learning has developed very fast over the past several years and has been used in many areas, including data analysis in physics \citep{baldi2014searching,adam2015higgs,sadowski2015deep}. Deep learning and conventional machine learning both offer ways to train models and classify data. Unlike the conventional machine learning method, which demands the user to extract relevant features from a classified object in order to train a model, the deep learning method skips the manual step of extracting features. Instead, one can feed the object directly into the deep learning algorithm which then predicts it. Deep learning is very powerful when dealing with complex objects, which is often the case in physics researches. We also expect an equally good, if not better, performance than conventional machine learning when deep learning is applied to DAMPE data for electron/proton separation.

We cannot use other detectors except for the BGO calorimeter to select pure electrons as a control sample or training sample from the on-orbit data for discriminator training like AMS-02. Our analysis relies strongly on consistency between the MC technique and real measurements.

\section{Conclusions}
\label{sect:conclusion}

Here the machine learning method is utilized to separate electrons from protons in DAMPE data from 2016 January 1 to 2017 June 30 in the energy range 10-100\,GeV. This machine learning method makes good use of all information available from the BGO calorimeter. It turns out that DAMPE can separate electrons from protons in energy range 10-100\,GeV with background rejection power of $9.0\times10^4$ using the machine learning method, which is higher compared to the background power rejection $>10^4$ at a signal efficiency 90\% provided by the traditional cut-based method used in \cite{ambrosi2017direct}.


\begin{acknowledgements}
This work was supported by State¨s Key Project of Research and Development Plan (2016YFA0400204), the National Natural Science Foundation of China (U1738133), Strategic Pioneer Research Program in Space Science of the Chinese Academy of Science (CAS), Youth Innovation Promotion Association of CAS, ministry of Science and Technology of Jiangsu Province (17KJD510001) and Changzhou Institute of Technology (YN1611).
\end{acknowledgements}



\bibliographystyle{raa}
\bibliography{ref}

\label{lastpage}

\end{document}